\documentclass[preprintnumbers,prd,twocolumn,tightenlines,floatfix,showpacs,amssymb,nofootinbib]{revtex4}

\usepackage{pictex}
\usepackage[dvips]{graphicx}
\usepackage{amsmath}

\usepackage{amsmath}
\usepackage{amsfonts}
\usepackage{amsthm}

\newcommand{\dd}{{\rm d}}
\newcommand{\del}{\partial}

\newcommand{\field}[1]{\mathbb{#1}}

\newcommand{\Z}{\field{Z}}

\DeclareMathOperator{\Tr}{Tr}
\DeclareMathOperator{\re}{Re}

\newcommand{\nn}{\nonumber}
\newcommand{\half}{\frac{1}{2}}
\newcommand{\third}{\frac{1}{3}}

\newcommand{\DD}{{\cal D}}
\newcommand{\expectation}[1]{\langle #1 \rangle}

\newcommand{\eqn}[1]{ \begin{equation} #1 \end{equation} }

\newcommand{\link}{
\setlength{\unitlength}{14pt}
\begin{picture}(1,1)(0,0)
\linethickness{0.25pt}
\put(0,0){\circle*{0.15}}
\put(0,0){\vector(1,0){1}}
\put(1,0){\circle{0.14}}
\end{picture}}
\newcommand{\stapleup}{
\setlength{\unitlength}{14pt}
\begin{picture}(1,1)(0,0)
\linethickness{0.25pt}
\put(0,0){\circle*{0.15}}
\put(0,0){\vector(0,1){1}}
\put(0,1){\vector(1,0){1}}
\put(1,1){\vector(0,-1){1}}
\put(1,0){\circle{0.14}}
\end{picture}}
\newcommand{\stapledown}{
\raisebox{-14pt}{
\setlength{\unitlength}{14pt}
\begin{picture}(1,1)(0,-1)
\linethickness{0.25pt}
\put(0,0){\circle*{0.15}}
\put(0,0){\vector(0,-1){1}}
\put(0,-1){\vector(1,0){1}}
\put(1,-1){\vector(0,1){1}}
\put(1,0){\circle{0.14}}
\end{picture}}}

\newcommand{\linkdag}{
\setlength{\unitlength}{14pt}
\begin{picture}(1,1)(0,0)
\linethickness{0.25pt}
\put(0,0){\circle*{0.15}}
\put(1,0){\vector(-1,0){1}}
\put(1,0){\circle{0.14}}
\end{picture}}
\newcommand{\stapleupdag}{
\setlength{\unitlength}{14pt}
\begin{picture}(1,1)(0,0)
\linethickness{0.25pt}
\put(0,0){\circle*{0.15}}
\put(0,1){\vector(0,-1){1}}
\put(1,1){\vector(-1,0){1}}
\put(1,0){\vector(0,1){1}}
\put(1,0){\circle{0.14}}
\end{picture}}
\newcommand{\stapledowndag}{
\raisebox{-14pt}{
\setlength{\unitlength}{14pt}
\begin{picture}(1,1)(0,-1)
\linethickness{0.25pt}
\put(0,0){\circle*{0.15}}
\put(0,-1){\vector(0,1){1}}
\put(1,-1){\vector(-1,0){1}}
\put(1,0){\vector(0,-1){1}}
\put(1,0){\circle{0.14}}
\end{picture}}}

\newcommand{\upx}{
\setlength{\unitlength}{14pt}
\begin{picture}(0.5,1)(0,0)
\linethickness{0.25pt}
\put(0.25,0){\circle*{0.15}}
\put(0.25,0){\vector(0,1){1}}
\end{picture}}

\newcommand{\downx}{
\raisebox{-14pt}{
\setlength{\unitlength}{14pt}
\begin{picture}(0.5,1)(0,-1)
\linethickness{0.25pt}
\put(0.25,0){\circle*{0.15}}
\put(0.25,0){\vector(0,-1){1}}
\end{picture}}}

\newcommand{\upxpmu}{
\raisebox{-14pt}{
\setlength{\unitlength}{14pt}
\begin{picture}(0.5,1)(0,-1)
\linethickness{0.25pt}
\put(0.25,0){\circle{0.14}}
\put(0.25,-1){\vector(0,1){1}}
\end{picture}}}

\newcommand{\downxpmu}{
\setlength{\unitlength}{14pt}
\begin{picture}(0.5,1)(0,0)
\linethickness{0.25pt}
\put(0.25,0){\circle{0.14}}
\put(0.25,1){\vector(0,-1){1}}
\end{picture}}

\newcommand{\upxdag}{
\setlength{\unitlength}{14pt}
\raisebox{-14pt}{
\begin{picture}(0.5,1)(0,-1)
\linethickness{0.25pt}
\put(0.25,0){\circle*{0.15}}
\put(0.25,-1){\vector(0,1){1}}
\end{picture}}}

\newcommand{\downxdag}{
\setlength{\unitlength}{14pt}
\begin{picture}(0.5,1)(0,0)
\linethickness{0.25pt}
\put(0.25,0){\circle*{0.15}}
\put(0.25,1){\vector(0,-1){1}}
\end{picture}}

\newcommand{\upxpmudag}{
\setlength{\unitlength}{14pt}
\begin{picture}(0.5,1)(0,0)
\linethickness{0.25pt}
\put(0.25,0){\circle{0.14}}
\put(0.25,0){\vector(0,1){1}}
\end{picture}}

\newcommand{\downxpmudag}{
\raisebox{-14pt}{
\setlength{\unitlength}{14pt}
\begin{picture}(0.5,1)(0,0)
\linethickness{0.25pt}
\put(0.25,1){\circle{0.14}}
\put(0.25,1){\vector(0,-1){1}}
\end{picture}}}

\newcommand{\rightdownxpmu}{
\setlength{\unitlength}{14pt}
\begin{picture}(1,1)(0,0)
\linethickness{0.25pt}
\put(1,0){\circle{0.14}}
\put(0,1){\vector(1,0){1}}
\put(1,1){\vector(0,-1){1}}
\end{picture}}

\newcommand{\downrightx}{
\raisebox{-14pt}{
\setlength{\unitlength}{14pt}
\begin{picture}(1,1)(0,-1)
\linethickness{0.25pt}
\put(0,0){\circle*{0.15}}
\put(0,0){\vector(0,-1){1}}
\put(0,-1){\vector(1,0){1}}
\end{picture}}}

\newcommand{\leftdownx}{
\setlength{\unitlength}{14pt}
\begin{picture}(1,1)(0,0)
\linethickness{0.25pt}
\put(0,0){\circle*{0.15}}
\put(1,1){\vector(-1,0){1}}
\put(0,1){\vector(0,-1){1}}
\end{picture}}

\newcommand{\downleftxpmu}{
\raisebox{-14pt}{
\setlength{\unitlength}{14pt}
\begin{picture}(1,1)(0,-1)
\linethickness{0.25pt}
\put(1,0){\circle{0.14}}
\put(1,0){\vector(0,-1){1}}
\put(1,-1){\vector(-1,0){1}}
\end{picture}}}

\newcommand{\leftupx}{
\setlength{\unitlength}{14pt}
\raisebox{-14pt}{
\begin{picture}(1,1)(-1,0)
\linethickness{0.25pt}
\put(-1,1){\circle*{0.15}}
\put(0,0){\vector(-1,0){1}}
\put(-1,0){\vector(0,1){1}}
\end{picture}}}

\newcommand{\upleftxpmu}{
\setlength{\unitlength}{14pt}
\begin{picture}(1,1)(0,0)
\linethickness{0.25pt}
\put(1,0){\circle{0.14}}
\put(1,0){\vector(0,1){1}}
\put(1,1){\vector(-1,0){1}}
\end{picture}}

\newcommand{\rectup}{
\setlength{\unitlength}{14pt}
\raisebox{-14pt}{
\begin{picture}(1,2)(0,0)
\linethickness{0.25pt}
\put(0,0){\circle*{0.15}}
\put(1,0){\vector(0,1){2}}
\put(1,2){\vector(-1,0){1}}
\put(0,2){\vector(0,-1){2}}
\put(1,0){\circle{0.14}}
\end{picture}}}

\newcommand{\rectdown}{
\setlength{\unitlength}{14pt}
\raisebox{-14pt}{
\begin{picture}(1,2)(0,-2)
\linethickness{0.25pt}
\put(0,0){\circle*{0.15}}
\put(1,0){\vector(0,-1){2}}
\put(1,-2){\vector(-1,0){1}}
\put(0,-2){\vector(0,1){2}}
\put(1,0){\circle{0.14}}
\end{picture}}}

\newcommand{\rectrightmidleft}{
\setlength{\unitlength}{14pt}
\begin{picture}(2,1)(0,0)
\linethickness{0.25pt}
\put(0,0){\circle*{0.15}}
\put(1,0){\vector(1,0){1}}
\put(2,0){\vector(0,1){1}}
\put(2,1){\vector(-1,0){2}}
\put(0,1){\vector(0,-1){1}}
\put(1,0){\circle{0.14}}
\end{picture}}

\newcommand{\rectrightmidright}{
\setlength{\unitlength}{14pt}
\begin{picture}(2,1)(0,0)
\linethickness{0.25pt}
\put(1,0){\circle*{0.15}}
\put(0,0){\vector(1,0){1}}
\put(2,0){\vector(0,1){1}}
\put(2,1){\vector(-1,0){2}}
\put(0,1){\vector(0,-1){1}}
\put(2,0){\circle{0.14}}
\end{picture}}

\newcommand{\rectleftmidleft}{
\setlength{\unitlength}{14pt}
\raisebox{-14pt}{
\begin{picture}(2,1)(0,-1)
\linethickness{0.25pt}
\put(0,0){\circle*{0.15}}
\put(1,0){\vector(1,0){1}}
\put(2,0){\vector(0,-1){1}}
\put(2,-1){\vector(-1,0){2}}
\put(0,-1){\vector(0,1){1}}
\put(1,0){\circle{0.14}}
\end{picture}}}

\newcommand{\rectleftmidright}{
\setlength{\unitlength}{14pt}
\raisebox{-14pt}{
\begin{picture}(2,1)(0,-1)
\linethickness{0.25pt}
\put(1,0){\circle*{0.15}}
\put(0,0){\vector(1,0){1}}
\put(2,0){\vector(0,-1){1}}
\put(2,-1){\vector(-1,0){2}}
\put(0,-1){\vector(0,1){1}}
\put(2,0){\circle{0.14}}
\end{picture}}}

\newcommand{\thinbf}{\fontseries{b}\selectfont}

\begin{document}
\preprint{ADP-04-04/T583}

\title{Hybrid Monte Carlo with Fat Link Fermion Actions}

\author{Waseem Kamleh}
\author{Derek B. Leinweber}
\author{Anthony G. Williams}
\affiliation{Special Research Centre for the Subatomic Structure of Matter and Department of Physics, University of Adelaide, 5005, Australia. }

\begin{abstract}

The use of APE smearing or other blocking techniques in lattice fermion actions can provide many advantages. There are many variants of these fat link actions in lattice QCD currently, such as FLIC fermions. The FLIC fermion formalism makes use of the APE blocking technique in combination with a projection of the blocked links back into the special unitary group. This reunitarisation is often performed using an iterative maximisation of a gauge invariant measure. This technique is not differentiable with respect to the gauge field and thus prevents the use of standard Hybrid Monte Carlo simulation algorithms. The use of an alternative projection technique circumvents this difficulty and allows the simulation of dynamical fat link fermions with standard HMC and its variants. The necessary equations of motion for FLIC fermions are derived, and some initial simulation results are presented. The technique is more general however, and is straightforwardly applicable to other smearing techniques or fat link actions.

\end{abstract}

\maketitle

Recent advances in computing power and in Lattice QCD (in particular, overlap fermions\cite{neuberger-massless}) have allowed simulations at sufficiently light quark masses to see that the behaviour of quenched QCD can differ from the true theory both qualitatively and quantitatively in the chiral regime\cite{zanotti03,young03}. As it is in the chiral regime where the difference from the quenched approximation will be highlighted, we would like to simulate at light quark masses in dynamical QCD. This is an extremely computationally expensive endeavour. Ideally, this would be done using overlap fermions, although large scale dynamical overlap simulations challenge the limits of current computing power, to say the least.

Fat Link Irrelevant Clover (FLIC) fermions have shown a number of promising advantages over standard actions, including improved convergence properties\cite{zanotti-hadron} and $O(a)$ improved scaling without the need for nonperturbative tuning\cite{excepcfg,zanotti-hadron2}. Furthermore, a reduced exceptional configuration problem has allowed efficient access to the light quark mass regime in the quenched approximation\cite{excepcfg}, where recent studies have highlighted deviations from the true theory\cite{zanotti03,young03}. As interest shifts ot focus on dynamical QCD, be it (truly) unquenched, or partially quenched, one might hope that the excellent behaviour at light quark mass displayed by FLIC fermions will carry over from the quenched theory to the unquenched one. This brings us to the issue of generating dynamical gauge field configurations with the fermionic determinant being that of the FLIC action. Brief accounts of this work were presented last year\cite{latt03,cairns03}. Recently, an alternative proposal for another type of smearing scheme that is differentiable has also appeared\cite{stout-links}.

\section{Hybrid Monte Carlo}

The standard technique for simulating dynamical fermions has for some time now been Hybrid Monte Carlo (HMC) \cite{hmc}. It is exact, ergodic and is $O(V^\frac{5}{4})$, that is, it scales almost linearly with the lattice volume $V.$ In order to introduce our notation and a framework for our technique, we briefly review the HMC algorithm for generating dynamical gauge field configurations.

We wish to generate an ensemble $\{ U_i \}$ of (statistically independent) representative gauge fields distributed according to the probability distribution
\eqn{ \rho(U_i) = e^{-S_{\rm eff}[U_i]}, }
where the effective action for full QCD
\begin{align}
S_{\rm eff}[U] &= S_{\rm g}[U] - \ln \det  D_{\rm f}[U], \\
\det  D_{\rm f} &= \int \DD \Bar{\psi} \DD \psi \ e^{- \int d^4x \Bar{\psi}(x) D_{\rm f} \psi(x)}
\end{align}
is obtained from the standard action 
\begin{align}
S[U,\Bar{\psi},\psi] &= S_{\rm g}[U] + S_{\rm f}[U,\Bar{\psi},\psi], \\
S_{\rm f} &= \int d^4x\ \Bar{\psi}(x) D_{\rm f}[U] \psi(x),
\end{align}
by integrating the fermionic degrees of freedom $\Bar{\psi},\psi$ of the functional integral
\eqn{ \expectation{{\cal O}} = \frac{1}{{\cal Z}} \int \DD U \DD \Bar{\psi} \DD \psi \ {\cal O}[U,\Bar{\psi},\psi]\ e^{-S[U,\Bar{\psi},\psi]}.}

For Wilson-like fermions in the physical region (away from exceptional configurations) $\det D_{\rm f}$ is real and positive. Hence if $M = D_{\rm f}^\dagger D_{\rm f},$ then $\det M = \det D_{\rm f}^2,$ and as $M$ is positive definite the following integral is convergent 
\eqn{ \frac{1}{\det M^{-1}} = \int \DD \phi^\dagger \DD \phi \ e^{- \int d^4x\ \phi^\dagger(x) M^{-1} \phi(x)},}
and as $\det M = \frac{1}{\det M^{-1}},$ we have the fermion determinant for two flavour QCD given in terms of an auxiliary pseudo-fermion field $\phi,$ so called because it is complex (bosonic) rather than Grassmannian. It is an identity that
\eqn{\det D_{\rm f} = \sqrt{\det D_{\rm f} \det D_{\rm f}^\dagger} = \sqrt{\det M} = \det \sqrt{M},}
and hence to simulate an odd number of sea quark flavours, it is possible to use $\det \sqrt{M}$\cite{oddhmc}.

For HMC, the four-dimensional lattice theory is made five-dimensional through the introduction of a fictitious (simulation) time, or evolution parameter $\tau.$ The gauge field $U$ is associated with its (fictitious) conjugate momenta $P$, and the five-dimensional system is described by the Hamiltonian,
\eqn{ {\cal H}[U,P] = \sum_{x,\mu}\ \half \Tr P_\mu(x)^2 + S_{\rm eff}[U].}
For Gaussian distributed $P$ the expectation value of an observable is unaffected by the 5D kinetic energy,
\begin{align}
\expectation{{\cal O}} &= \frac{1}{\cal Z}\int \DD P \DD U\ {\cal O}[U] e^{- {\cal H}[U,P]},\\
{\cal Z} &= \int \DD P \DD U e^{- {\cal H}[U,P]}.
\end{align}
Given $U,$ a new gauge field $U'$ is generated by the update $U \to U',$ which consists of
\begin{enumerate}
\item[(i)] \emph{Refreshment:} Sample $P$ from a Gaussian ensemble, $\rho(P) \propto e^{-\frac{1}{2}\Tr P^2}.$  Generate a pseudo-fermionic background field $\phi$ according to $\rho(\phi) \propto e^{-\phi^\dagger M^{-1} \phi}.$

\item[(ii)] \emph{Molecular Dynamics Trajectory:} Integrate Hamilton's equations of motion to deterministically evolve $(U,P)$ along a phase space trajectory to $(U',P').$

\item[(iii)] \emph{Metropolis step:} Accept or reject the new configuration $(P',U')$ with probability $\rho(U \to U') = \min(1,e^{-\Delta {\cal H}}), \Delta {\cal H} = {\cal H}[U'] - {\cal H}[U].$
\end{enumerate}

The discretised equations of motion are derived by requiring that the Hamiltonian be conserved along the trajectory, $\frac{\dd{\cal H}}{\dd\tau} = 0.$ The following discretised equations of motion then approximately conserve ${\cal H}$ for small step sizes, $\Delta \tau,$
\begin{align}
\label{eq:Umuupdate} U_\mu(x,\tau + \Delta\tau) &= U_\mu(x,\tau)\exp\big(i\Delta\tau P_\mu(x,\tau)\big), \\
\label{eq:Pmuupdate} P_\mu(x,\tau + \Delta\tau) &= P_\mu(x,\tau) - U_\mu(x,\tau)\frac{\delta S_{\rm eff}}{\delta U_\mu(x,\tau)}. 
\end{align}
In our implementation, we evaluate the matrix exponential directly through diagonalisation, rather than expanding it.

FLIC fermions\cite{zanotti-hadron} are clover-improved fermions where the irrelevant operators are constructed using APE smeared links \cite{ape-one,ape-two}. As with other efficient updating algorithms, HMC makes use of the variation of the action with respect to the links $\frac{\delta S}{\delta U}$ in order that the proposed configurations have high acceptance rates. Previously, it has not been clear how to perform HMC with fermion actions that make use of the APE blocking technique in combination with a projection of the blocked links back into the special unitary group. This reunitarisation is often performed using an iterative maximisation of a gauge invariant measure, and this choice of reuniterisation is the source of the difficulty. The problem is that the iterative technique is not differentiable with respect to the gauge field and thus it is not possible to calculate $\frac{\delta S}{\delta U},$ which is necessary for the equations of motion above. In the next section we consider an alternative technique and show that is does not suffer from this problem, allowing the simulation of dynamical fat link fermions with standard HMC (and its variants). 

\section{SU(3) Projection}
\label{sec:projectsu3}

The APE smeared links $U^{(n)}_\mu(x)$ present in the FLIC fermion action are constructed from $U_\mu(x)$ by performing $n$ smearing sweeps, where in each sweep we first perform an APE blocking step,
\begin{equation}
\label{eq:apesmear}V^{(j)}_\mu(x)[U^{(j-1)}] = (1-\alpha)\ \link + \frac{\alpha}{6} \sum_{\nu \ne \mu}\ \stapleup\ + \stapledown\ , 
\end{equation}
followed by a projection back into $SU(3), U^{(j)}_\mu(x) = {\mathcal P}(V^{(j)}_\mu(x)).$ Frequently, the projection is performed using an algorithm which updates $U^{(j)}$ iteratively in order to maximise the following gauge invariant measure,
\begin{equation}
U^{(j)}_\mu(x) \in \{ U' \in SU(3) | \re \Tr (U' V^{(j)\dagger}_\mu(x)) \text{ is maximal} \}.
\end{equation}
We refer to this projection technique as MaxReTr projection. While this projection minimises the local action\cite{derek-impcooling}, as we mentioned earlier it is not differentiable with respect to $U_\mu(x)$ and hence not suitable for use in HMC.

Now, given any matrix $X$, then $X^\dagger X$ is hermitian and may be diagonalised. Then it is possible (for $\det X \ne 0$) to define a matrix 
\begin{equation}
W = X\frac{1}{\sqrt{X^\dagger X}}
\end{equation}
whose spectrum lies on the complex unit circle and is hence unitary ($w(z)=z/z^*z$ is the complex version of the sign function). Furthermore, $W$ possesses the same gauge transformation properties as $X.$ This is easily seen. Let $X_\mu(x)$ transform as
\eqn{X_\mu(x) \to G(x)X_\mu(x)G^\dagger(x+e_\mu),}
then 
\eqn{X_\mu^\dagger(x) \to G(x+e_\mu) X_\mu^\dagger(x) G^\dagger(x),}
and hence 
\eqn{X_\mu^\dagger(x) X_\mu(x) \to G(x+e_\mu) X_\mu^\dagger(x) X_\mu(x) G^\dagger(x+e_\mu).}
Noting that $[X_\mu^\dagger(x) X_\mu(x)]^{-\half}$ has the same transformation properties as $X_\mu^\dagger(x) X_\mu(x)$ it is then straight forward to see that 
\eqn{W_\mu(x) \to G(x)W_\mu(x)G^\dagger(x+e_\mu),}
as required.

Given the unitary matrix $W,$ we can construct another matrix, 
\begin{equation}
W' = \frac{1}{\sqrt[3]{\det W}} W
\end{equation}
which is special unitary. Earlier work\cite{liu-projection} has incorrectly omitted the cube root. As there are three different complex roots, we have a $\Z_3$ ambiguity which we break by choosing the principal value of the cube root\footnote{For complex $z,$ the principal value of the cube root satisfies $-\frac{\pi}{3} < \arg \sqrt[3]{z} < \frac{\pi}{3}.$ For purely real $z,$ we choose $\sqrt[3]{z}$ to be real.} In selecting the principal value, the projected matrices lie closest to those given by the MaxReTr method, and are hence smoother. The mean plaquette is closer to unity thus minimising the action. We refer to this technique for projecting $X_\mu(x)$ into the special unitary group as unit circle projection. 

The two methods produce smeared links that are different but numerically close (according to the usual matrix norm, $||A|| = \sqrt{\lambda_{\rm max}(A^\dagger A)}$). Using the mean link as a measure of the smoothness of the smeared gauge field, Table \ref{tab:uzero} indicates that the two methods presented here produce equally smooth gauge fields. 

\begin{table}[t]
\begin{center}
\begin{tabular}{ccc}
Sweep & Unit Circle & MaxReTr \\
\hline
  0   & {\thinbf 0.866138301214314}   & {\thinbf 0.866138301214314}\\
  1   & {\thinbf 0.9603}13394813806   & {\thinbf 0.9603}48747275940 \\
  2   & {\thinbf 0.9807}35000838119   & {\thinbf 0.9807}51346847750 \\
  3   & {\thinbf 0.9883}84926461589   & {\thinbf 0.9883}93707639555 \\
  4   & {\thinbf 0.99210}3013943516   & {\thinbf 0.99210}7844842705 \\
  5   & {\thinbf 0.99418}2852413813   & {\thinbf 0.99418}5532052157 \\ 
  6   & {\thinbf 0.99545}7365275018   & {\thinbf 0.99545}8835653863 \\
  7   & {\thinbf 0.99629}3668622924   & {\thinbf 0.99629}4454083006 \\
  8   & {\thinbf 0.996878}305318083   & {\thinbf 0.996878}710433084 
\end{tabular}
\end{center}
\caption{\label{tab:uzero} The mean link $u_0$ for a single configuration as a function of number of APE smearing sweeps at $\alpha=0.7$, for the two different projection methods. The boldface indicates significant digits which match. The configuration is a dynamical gauge field with DBW2 glue and FLIC sea fermions, at $\beta = 8.0,\kappa = 0.1280\ (a=0.17 \text{ fm},m_{\rm q}=200\text{ MeV}).$}
\end{table}

 While numerically the two methods may be nearly equivalent, unit circle projection possesses a significant advantage over MaxReTr projection. The matrix inverse square root function can be approximated by a rational polynomial (whose poles lie on the imaginary axis)\cite{neuberger-practical,Chiu-zolotarev}, $W[X] \approx W_k[X],$
\begin{equation}
W_k[X] \equiv d_0 X (X^\dagger X + c_{2n})\sum_{l=1}^{k} \frac{b_l}{X^\dagger X + c_l},
\end{equation}
where the formula for the coefficients $d_0,b_l,c_l$ can be found Ref.~\cite{Chiu-zolotarev}. This approximation is differentiable in a matrix sense for all $X$ for which the inverse square root can be defined. This means that we can construct $\frac{\delta S}{\delta U}$ for fermion actions which involve unit circle projection, and hence it is a reuniterisation method which is compatible with HMC.

\section{Equations of Motion}

Having now defined the APE smearing prescription (with projection) in a differentiable closed form, we proceed to derive the equations of motion necessary for the use of the HMC algorithm with FLIC fermions.

\subsection{Mathematical Preliminaries}

The equations of motion are derived using multi-variate calculus. To make the derivation simple and provide an understanding of how best to implement the equations efficiently, we develop some appropriate mathematical tools. Using index notation, we define a (minimal) set of tensor operations (including differentiation) such that we can perform the derivation in an index free language. 

The derivative of a real-valued function $f[A]$ with respect to the matrix $A$ is a rank 2 type (1,1) tensor (distinguishing contravariant and covariant indices),
\eqn{ [\frac{\del f}{\del A}]^i{}_j = \frac{\del}{\del A^j{}_i} f[A].} 
The derivative of a matrix-valued function $M[A]$ with respect to the matrix $A$ is a rank 4 type (2,2) tensor,
\eqn{ [\frac{\del M}{\del A}]^i{}_j{}^k{}_l = \frac{\del}{\del A^j{}_k} M[A]^i{}_l.}
The set of type $(m,n)$ tensors ${\cal T}{}^m_n$ forms a vector space. We define the outer product $\otimes : {\cal T}{}^1_1 \times {\cal T}{}^1_1 \to {\cal T}{}^2_2$ as
\eqn{ (A \otimes B)^i{}_j{}^k{}_l = A^i{}_j B^k{}_l.}
Noting carefully the index ordering, define the ``direct'' product $\oplus : {\cal T}{}^1_1 \times {\cal T}{}^1_1 \to {\cal T}{}^2_2$ as
\eqn{ (A \oplus B)^i{}_j{}^k{}_l = A^k{}_j B^i{}_l.}
Given a scalar function $f[B]$ and a matrix function $B[A]$ the (scalar-matrix) chain rule states
\eqn{ \frac{\del f}{\del A} = \frac{\del f}{\del B} \star \frac{\del B}{\del A},}
where we define the contraction induced by the chain rule as the (rank 2) star product, $\star : {\cal T}{}^1_1 \times {\cal T}{}^2_2 \to {\cal T}{}^1_1,$ with 
\eqn{(A \star T)^i{}_l = A^j{}_k T^i{}_j{}^k{}_l.}
Given two matrix functions $M[B]$ and $B[A]$, the  (matrix-matrix) chain rule states
\eqn{\frac{\del M}{\del A} = \frac{\del M}{\del B} \star \frac{\del B}{\del A},}
where we define the contraction induced by this chain rule as the (rank 4) star product, $\star : {\cal T}{}^2_2 \times {\cal T}{}^2_2 \to {\cal T}{}^2_2,$ with
\eqn{ (S \star T)^i{}_j{}^k{}_l = S^i{}_m{}^n{}_l T^m{}_j{}^k{}_n. }
It is interesting to note that the star product induces an algebra structure on the vector space of type (2,2) tensors, that is, $({\cal T}{}^2_2,+,\star)$ is an algebra with multiplicative identity $I \otimes I$.

We define juxtaposition for $A \in {\cal T}{}^1_1, T \in {\cal T}{}^2_2$ by the contractions
\begin{align}
(AT)^i{}_j{}^k{}_l &= A^i{}_m T^m{}_j{}^k{}_l,\\
(TA)^i{}_j{}^k{}_l &= T^i{}_j{}^k{}_m A^m{}_l.
\end{align}

All our derivatives will be derived from the basic matrix differentiation rule. Given matrices $M,A,B,C$ then for $M = ABC$ we have
\eqn{ \frac{\del M}{\del B} = A \otimes C. }
An immediate consequence of this is that
\eqn{ \frac{\del M}{\del M} = I \otimes I.}
The (scalar-matrix) product rule is 
\eqn{\frac{\del}{\del A}(fM) = \frac{\del f}{\del A}\oplus M + f\frac{\del M}{\del A}.}
The (matrix-matrix) product rule is
\begin{align}
\label{eq:matmatprod}\frac{\del}{\del A}(XY) = X\frac{\del Y}{\del A} + \frac{\del X}{\del A}Y,
\end{align}
which is easily shown to imply the identity
\eqn{\label{eq:matinverse}\frac{\del X^{-1}}{\del A} = -X^{-1} \frac{\del X}{\del A} X^{-1}.}
In the following sections we will make use of the identity
\eqn{ A \star (B \oplus C) = (A\cdot B) C,}
where $A\cdot B = A^i{}_jB^j{}_i.$ Additionally, of particular numerical importance is the identity
\eqn{ A \star (B \otimes C) = BAC, }
which has two major benefits. It allows us to evaluate two matrix multiplications instead of an outer product (computational saving), hence enabling us to implement the equations of motion without having to store any tensor fields (memory saving).

\subsection{Standard Derivatives}

The equations of motion for FLIC fermions are derived starting from the equations for the standard clover fermion action. We divide the effective action into its gauge part and pseudo-fermionic part,
\eqn{ S_{\rm eff} = S_{\rm g} + S_{\rm pf}.}
We reformulate some standard results in terms of the mathematics of the previous section. We will adopt a more convenient notation for quantities with a lattice site index  $x,$ using a subscript $U_{\mu,x}$ rather than $U_\mu(x).$ The matrix products of link variables are often denoted diagrammatically. 
\begin{widetext}
For a plaquette plus rectangle improved gauge action,
\eqn{ S_{\rm g} = \sum_{x,\mu < \nu} \re \Tr \left[ \beta_{1\times 1} (1 - U_{\mu\nu}(x)) + \beta_{2\times 1} (1 - R^{2\times 1}_{\mu\nu}(x)) + \beta_{1\times 2} (1 - R^{1\times 2}_{\mu\nu}(x)) \right],}
we have
\begin{multline}
\frac{\del S_{\rm g}}{\del U_{\mu,x}} = - \sum_{\nu \ne \mu} \beta_{1\times 1} \left( \stapleupdag + \stapledowndag \right) + \beta_{1\times 2} \left(  \rectup + \rectdown\ \right) + \beta_{2\times 1} \left( \rectleftmidleft + \rectleftmidright + \rectrightmidleft + \rectrightmidright\ \right),
\end{multline}
where the filled circles indicate the point $x.$ The coefficients $\beta_{m\times n} = \third \beta c_{m\times n}$ depend on the choice of gauge action. For L\"uscher-Weisz glue, $c_{1\times1} = \frac{5}{3}, c_{2\times1}=c_{1\times2}=-\frac{1}{12}.$ For DBW2 glue, we choose coefficients that are normalised such that $c_{1\times1}=1.$

The pseudo-fermionic action is $S_{\rm pf} = - \sum_x \phi^\dagger_x \eta_x$, where $\eta = (D^\dagger D)^{-1} \phi$, hence by equations (\ref{eq:matmatprod}) and (\ref{eq:matinverse}) we have
\eqn{ \frac{\del S_{\rm pf}}{\del U_{\mu,x}} = \phi^\dagger (D^\dagger D)^{-1} \left( D^\dagger\frac{\del D}{\del U_{\mu,x}} + \frac{\del D^\dagger}{\del U_{\mu,x}} D \right) (D^\dagger D)^{-1} \phi.}
Setting $\chi = D\eta$, we obtain
\eqn{ \frac{\del S_{\rm pf}}{\del U_{\mu,x}} = \chi^\dagger \frac{\del D}{\del U_{\mu,x}}\eta + \eta^\dagger\frac{\del D^\dagger}{\del U_{\mu,x}} \chi.}
Now, the FLIC action is explicitly given by 
\begin{multline}
(D_{\rm flic}\psi)_x = -\half \sum_\mu \big(\frac{U^{\rm fl}_{\mu,x}}{u_0^{\rm fl}}-\gamma_\mu \frac{U_{\mu,x}}{u_0}\big)\psi_{x+\mu} + \big(\frac{U^{{\rm fl}\dagger}_{\mu,x}}{u_0^{\rm fl}}+\gamma_\mu \frac{U^\dagger_{\mu,x}}{u_0}\big)\psi_{x-\mu} 
+ (4+m - \frac{1}{4 u_0^{\rm fl 4}}\sigma_{\mu\nu}F^{\rm cl}_{\mu\nu,x})\psi_x 
\end{multline}
and contains three terms, the Dirac term (constructed with standard links), the Wilson term and the clover term (using fat links for $F^{\rm cl}_{\mu\nu,x},$ and setting $\sigma_{\mu\nu} = \half[\gamma_\mu,\gamma_\nu]$). Hence we may decompose the pseudofermionic derivative into three terms also. The first comes from the Dirac term,
\eqn{ \frac{\del S^{\rm d}_{\rm pf}}{\del U_{\mu,x}} = \frac{1}{2u_0}\Tr_{\rm spin} (\eta^\dagger_x \otimes \gamma_\mu \chi_{x+\mu} + \chi^\dagger_x \otimes \gamma_\mu \eta_{x+\mu}),}
while the Wilson and clover terms only explicit dependence is on the smeared links,
\begin{align}
\frac{\del S^{\rm w}_{\rm pf}}{\del U^{\rm fl}_{\mu,x}} &= -\frac{1}{2u_0^{\rm fl}}\Tr_{\rm spin} (\eta^\dagger_x \otimes \chi_{x+\mu} + \chi^\dagger_x \otimes \eta_{x+\mu}) \\
\frac{\del S^{\rm cl}_{\rm pf}}{\del U^{\rm fl}_{\mu,x}}&= -\frac{1}{4u_0^{\rm fl 4}} \Tr_{\rm spin} (\eta^\dagger_y \sigma_{\nu\lambda} \frac{\del F_{\nu\lambda,y}}{\del U^{\rm fl}_{\mu,x}} \chi_y + \chi^\dagger_y \sigma_{\nu\lambda} \frac{\del F_{\nu\lambda,y}}{\del U^{\rm fl}_{\mu,x}} \eta_y),
\end{align}
where the vector outer product defines a matrix $(\eta \otimes \chi^\dagger)^i{}_j = \eta^i\chi_j^*.$ The one loop clover term is given by $F_{\nu\lambda,y} = \frac{1}{8} ( C_{\nu\lambda,y} - C_{\nu\lambda,y}^\dagger),$ where 
\eqn{C_{\nu\lambda,y} = U_{(+\nu)(+\lambda),y} + U_{(+\lambda)(-\nu),y} + U_{(-\lambda)(+\nu),y} + U_{(-\nu)(-\lambda),y},}
and $U_{(\pm\nu)(\pm\lambda),y}$ indicates the plaquette starting at $y,$ oriented in the $\nu-\lambda$ plane, with the first (second) link in the direction indicated by the first (second) index. When differentiating with respect to $U^{\rm fl}_{\mu,x},$ any terms where $y$ lies further away from $x$ than $x\pm\nu\pm\lambda$ will be zero. Further, noting that the derivative is zero unless either $\nu=\mu$ or $\lambda=\mu$ and $\nu\ne\lambda$ we can without loss of generality choose $\mu=\nu.$ Letting $\mu=\nu$ be in the horizontal direction and $\lambda$ be in the transverse (vertical) direction, the contribution to the derivative due to the clover term is
\begin{multline}
\frac{\del F_{\nu\lambda,y}}{\del U^{\rm fl}_{\mu,x}} = \frac{1}{8} \left( I \otimes \stapleupdag\ \delta^{\nu\mu}_{y,x} + \downxdag \otimes \upleftxpmu\  \delta^{\nu\mu}_{y,x+\lambda} - I \otimes \stapledowndag\ \delta^{\nu\mu}_{y,x} - \upxdag \otimes \downleftxpmu\ \delta^{\nu\mu}_{y,x-\lambda} \right. \\
\left. + \stapleupdag \otimes I\ \delta^{\nu\mu}_{y,x+\mu} + \leftdownx \otimes \upxpmudag\ \delta^{\nu\mu}_{y,x+\mu+\lambda} - \leftupx \otimes \downxpmudag\ \delta^{\nu\mu}_{y,x+\mu-\lambda} - \stapledowndag \otimes I\ \delta^{\nu\mu}_{y,x+\mu} \right),
\end{multline}
where the filled circles indicate the point $x,$ and the point $y$ is located at the start (end) of the diagrams that lie on the left (right) side of the outer product, as can be deduced from the Kronecker $\delta$-s.

\end{widetext}

\subsection{Smeared Link Derivatives}

 Now, having constructed the explicit derivatives of $S_{\rm pf}$ with respect to the thin and fat links, the total derivative with respect to the thin links is
\eqn{ \frac{\dd S_{\rm pf}}{\dd U_{\mu,x}} = \frac{\del S_{\rm pf}}{\del U_{\mu,x}} + \frac{\del S_{\rm pf}}{\del U^{\rm fl}_{\nu,y}} \star \frac{\dd U^{\rm fl}_{\nu,y}}{\dd U_{\mu,x}}.}
If we have performed $n$ sweeps of APE smearing to form the fat links, then the right hand term is constructed through $n$ applications of the chain rule
\eqn{ \frac{\dd S}{\dd U^{(j-1)}_{\mu,x}} = \frac{\del S}{\del U^{(j)}_{\nu,y}} \star \frac{\del U^{(j)}_{\nu,y}}{\del U^{(j-1)}_{\mu,x}} + \frac{\del S}{\del U^{(j)\dagger}_{\nu,y}} \star \frac{\del U^{(j)\dagger}_{\nu,y}}{\del U^{(j-1)}_{\mu,x}}, }
until we arrive at $\frac{\dd S}{\dd U^{(0)}_{\mu,x}}.$ We note here that the partial derivative with respect to a forward (backward) link only picks up terms that contain the forward (backward) link, and not its conjugate (that is, $U$ and $U^\dagger$ are considered independent with regard to partial differentiation, see Eqs. (\ref{eq:apesmear}),(\ref{eq:dVbydU}),(\ref{eq:dVdagbydU})).  For the sake of both computational efficiency and simplicity, this chain rule is itself composed of several chain rules, and hence evaluated in several steps. Each step corresponds to a step in the APE smearing process, but we go through them in reverse order. 

The final step in the APE smearing process (with unit circle projection) is
\eqn{U^{(n)}_{\mu,x} = \frac{1}{\det W^{(n)}_{\mu,x}{}^{-\third}} W^{(n)}_{\mu,x}.}
Therefore the first chain rule corresponds to this step,
\begin{align}
\nn \frac{\del S}{\del W_{\mu,x}} &= \frac{\del S}{\del U^{(n)}_{\mu,x}} \star \frac{\del U^{(n)}_{\mu,x}}{\del W_{\mu,x}} \\
\nn &= \frac{\del S}{\del U^{(n)}_{\mu,x}} \star \Big( -\third \det W_{\mu,x}^{-\frac{4}{3}} \frac{\del \det W_{\mu,x}}{\del W_{\mu,x}} \oplus W_{\mu,x} \\ &\qquad + \det W_{\mu,x}^{-\frac{1}{3}} I \otimes I \Big),
\end{align}
where
\eqn{ \det A = \epsilon^{ijk} A^1{}_i A^2{}_j A^3_k }
and hence denoting the permutations $\pi_i = (i \bmod 3) + 1, \pi^2_i = \pi_{\pi_i},$
\eqn{ \frac{\del \det A}{\del A^i{}_j} = \epsilon_{jlm} A^{\pi_i}{}_l A^{\pi^2_i}{}_m.}
There are several chain rules that correspond to 
\eqn{W^{(n)}_{\mu,x} = V^{(n)}_{\mu,x}(V^{(n)\dagger}_{\mu,x} V^{(n)}_{\mu,x})^{-\half}.}
For the first, we define $H_{\mu,x} = V^\dagger_{\mu,x} V_{\mu,x}$. Then
\eqn{ \frac{\del S}{\del H_{\mu,x}} = \frac{\del S}{\del W_{\mu,x}} \star \frac{\del W_{\mu,x}}{\del H_{\mu,x}}.}
Using
\eqn{ W_{\mu,x} \approx d_0 V_{\mu,x} (H_{\mu,x} + c_0)\sum_{l=1}^{k} \frac{b_l}{H_{\mu,x} + c_l},}
we have
\begin{multline}
\frac{\del W_{\mu,x}}{\del H_{\mu,x}} = d_0 V_{\mu,x} \left( I \otimes \sum_{l=1}^{k} \frac{b_l}{H_{\mu,x} + c_l} \right. \\
\left. -  (H_{\mu,x} + c_0)\sum_{l=1}^{k} b_l \frac{1}{H_{\mu,x} + c_l} \otimes \frac{1}{H_{\mu,x} + c_l}\right).
\end{multline}
We can then construct
\begin{align}
\nn \frac{\del S}{\del V_{\mu,x}} &= \frac{\del S}{\del W_{\mu,x}} \star \frac{\del W_{\mu,x}}{\del V_{\mu,x}} + \frac{\del S}{\del H_{\mu,x}} \star \frac{\del H_{\mu,x}}{\del V_{\mu,x}} \\
 &= \frac{\del S}{\del W_{\mu,x}} \star (I \otimes  H_{\mu,x}^{-\half}) + \frac{\del S}{\del H_{\mu,x}} \star (V^\dagger_{\mu,x} \otimes I),
\end{align}
and also
\begin{align}
\nn \frac{\del S}{\del V^\dagger_{\mu,x}} &= \frac{\del S}{\del W^\dagger_{\mu,x}} \star \frac{\del W^\dagger_{\mu,x}}{\del V^\dagger_{\mu,x}} + \frac{\del S}{\del H_{\mu,x}} \star \frac{\del H_{\mu,x}}{\del V^\dagger_{\mu,x}} \\
&= \frac{\del S}{\del W^\dagger_{\mu,x}} \star (H_{\mu,x}^{-\half} \otimes I) + \frac{\del S}{\del H_{\mu,x}} \star (I \otimes  V_{\mu,x}).
\end{align}
Last, we make use of the chain rule
\eqn{ \frac{\del S}{\del U^{(n-1)}_{\mu,x}} = \frac{\del S}{\del V_{\nu,y}}\star \frac{\del V_{\nu,y}}{\del U^{(n-1)}_{\mu,x}} + \frac{\del S}{\del V^\dagger_{\nu,y}}\star \frac{\del V^\dagger_{\nu,y}}{\del U^{(n-1)}_{\mu,x}},}
where
\begin{align}
V_{\nu,y} &= (1-\alpha)\ \link + \frac{\alpha}{6} \sum_{\lambda \ne \nu}\ \stapleup\ + \stapledown\ ,\\
V^\dagger_{\nu,y} &= (1-\alpha)\ \linkdag + \frac{\alpha}{6} \sum_{\lambda \ne \nu}\ \stapleupdag\ + \stapledowndag\ ,
\end{align}
and $y$ is indicated by the filled circle.
\begin{widetext}
It is then straightforward to show that
\begin{multline}
\label{eq:dVbydU}\frac{\del V_{\nu,y}}{\del U^{(n-1)}_{\mu,x}} = (1-\alpha) I \otimes I + \frac{\alpha}{6} \sum_{\lambda \ne \nu}\ \upx \otimes \downxpmu\ \delta^{\mu\nu}_{y,x-\lambda}
+ \downx \otimes \upxpmu\ \delta^{\mu\nu}_{y,x+\lambda} + I \otimes \rightdownxpmu\ \delta^{\mu\lambda}_{y,x} + \downrightx \otimes I\ \delta^{\mu\lambda}_{y,x+\lambda-\nu}.
\end{multline}
and
\eqn{\label{eq:dVdagbydU} \frac{\del V^\dagger_{\nu,y}}{\del U^{(n-1)}_{\mu,x}} = \frac{\alpha}{6} \sum_{\lambda \ne \nu}\ I \otimes \leftdownx\ \delta^{\mu\lambda}_{y,x-\nu} + \downleftxpmu\ \otimes I\ \delta^{\mu\lambda}_{y,x+\lambda},}
where in these diagrams the filled circle indicates the point $y.$
Hence,
\begin{multline}
\frac{\del S}{\del U^{(n-1)}_{\mu,x}} = (1-\alpha) \frac{\del S}{\del V_{\mu,x}} + \frac{\alpha}{6} \sum_{\nu \ne \mu}\ \frac{\del S}{\del V_{\mu,x-\nu}} \star \upx \otimes \downxpmu\ + \frac{\del S}{\del V_{\mu,x+\nu}} \star \downx \otimes \upxpmu \\ + \frac{\del S}{\del V_{\nu,x}} \star I \otimes \rightdownxpmu\ + \frac{\del S}{\del V_{\nu,x+\mu-\nu}} \star \downrightx \otimes I 
+ \frac{\del S}{\del V^\dagger_{\mu,x-\nu}} \star I \otimes \leftdownx\  + \frac{\del S}{\del V^\dagger_{\mu,x+\mu}} \star \downleftxpmu\ \otimes I,
\end{multline}
where now $x$ is indicated by the filled circle.

Having constructed the total derivative of the action with respect to $U_{\mu,x},$ we can calculate the variation of $S$ with respect to the gauge field (noting $U^\dagger = U^{-1}$),
\eqn{\frac{\delta S}{\delta U} = \frac{\dd S}{\dd U} + \frac{\dd S}{\dd U^\dagger} \star \frac{\del U^\dagger}{\del U} = \frac{\dd S}{\dd U} - U^\dagger \frac{\dd S}{\dd U^\dagger} U^\dagger,}
and hence the necessary equations of motion, (\ref{eq:Umuupdate}) and (\ref{eq:Pmuupdate}). It is numerically efficient to make use of the fact that 
\eqn{\frac{\dd S}{\dd U^\dagger} = \left( \frac{\dd S}{\dd U} \right)^\dagger.} 
\end{widetext}

\section{Simulation Results}

We have implemented the equations above within a standard two-flavour HMC, with multiple time scales. Expensive pseudo-fermion momenta updates are performed at a larger step size $\Delta\tau = \Delta\tau_{\rm pf}$ and the cheaper gauge momenta updates are performed more often, $\Delta\tau_{\rm g} = \frac{1}{n} \Delta\tau_{\rm pf},$ for some integer $n.$ Molecular dynamics trajectories are of unit length, $n_{\rm md} \Delta\tau = 1.$ In particular, we have implemented a modified version of the Ritz algorithm to diagonalise arrays of $3\times3$ matrices in parallel. This routine is used in the SU(3) projection step, and is also used to calculate the matrix exponentials that are needed in other parts of the algorithm, avoiding the need to use polynomial approximations to the exponential. This means that the accuracy of the exponential in Eq. (\ref{eq:Umuupdate}) does not depend upon the step size $\Delta \tau.$

An eighth order Zolotarev approximation to the inverse square root is used to approximate $W_{\mu,x}$ in unit circle projection. We find that the spectral range at this order is ample. In smooth gauge backgrounds it is easily shown that unit circle projection is well defined, that is $\det V^\dagger_{\mu,x}V_{\mu,x} > 0.$

If we assume a smoothness condition $||1-U_{\mu\nu,x}|| \le \epsilon\; \forall x,\mu,\nu,$ then we have a lower bound
\eqn{ \nn V^\dagger_{\mu,x}V_{\mu,x} \ge 1 - 2 \alpha\epsilon - \alpha^2\epsilon^2. } 
To prove this, we note that APE blocking may be written in terms of the plaquette field,
\eqn{ V_{\mu,x} = U_{\mu,x}\big(1 - \frac{\alpha}{6}\sum_{\pm\nu\ne\mu} (1 - U_{\mu\nu,x}^\dagger)\big). }
Define $Z = \sum_{\pm\nu\ne\mu} (1 - U_{\mu\nu,x}^\dagger),$ then
\eqn{ V^\dagger_{\mu,x}V_{\mu,x} = 1 - \frac{\alpha}{6}(Z + Z^\dagger) + \frac{\alpha^2}{36} Z^\dagger Z. }
As $||Z||\le 6 \epsilon,$ we then have
\eqn{ \lambda_{\rm min}(V^\dagger_{\mu,x}V_{\mu,x}) \ge 1 - 2 \alpha\epsilon - \alpha^2\epsilon^2, } 
which is strictly positive for small enough $\epsilon$.

While the smeared link equations of motions are complex, our implementation evaluates them efficiently due to the optimisations that can be performed through the calculus we constructed earlier. At large sea quark masses the code already spends over 90\% of its time in the CG inversion required to calculate $\eta = (D^\dagger D)^{-1}\phi,$ and as the quark mass decreases this fraction increases. So as is standard, the generation of dynamical gauge fields is dominated by the CG inversion. Simulation results are presented in Table \ref{tab:hmcresults}.

It is a simple exercise to apply our results to generate gauge fields with dynamical FLIC Overlap quarks, although this would be extremely computationally intensive. The availability of HMC as a simulation algorithm for dynamical FLIC fermions is significant, as it scales almost linearly with the lattice volume $V$, whereas previously there were only $O(V^2)$ alternatives \cite{hasenfratz-dynamical}. Furthermore, the method we have described is general and can be straightforwardly applied to any fermion action with reuniterisation, including overlap fermions with a fat link kernel \cite{kamleh-overlap,degrand-fatover,bietenholz,kovacs}, or other types of fat link actions \cite{stephenson} that may involve alternative smearing techniques \cite{hasenfratz-hyp}. Additionally, any of the variants of HMC can be also be used, in particular Polynomial HMC\cite{polynomial-hmc} or Rational HMC\cite{rational-hmc} which allow for the simulation of odd numbers of sea quark flavours.

\begin{acknowledgments}
The authors would like to thank Herbert Neuberger for valuable correspondence. This work was supported by the Australian Research Council (ARC). Computations were performed on facilities provided by the Australian Partnership for Advanced Computing (APAC), the South Australian Partnership for Advanced Computing (SAPAC) and the National Computing Facility for Lattice Gauge Theory (NCFLGT).
\end{acknowledgments}


\begin{table*}[p]
\begin{center}
\begin{tabular}{ccccccccc}
$\beta$ & $\kappa$ & $S_{\rm gauge}$ & $\Delta\tau$ & $\frac{\Delta\tau_{\rm pf}}{\Delta\tau_{\rm g}}$ & $\rho_{\rm acc}$ & $u_0$ & $a$ & $m_\pi$ \\
\hline
3.6 & 0.1347 & IMP  & 0.0143 & 2 & 0.55 & 0.8226 & 0.247(9) & 0.702 \\
3.7 & 0.1340 & IMP  & 0.0147 & 2 & 0.64 & 0.8338 & 0.218(4) & 0.680 \\
3.8 & 0.1332 & IMP  & 0.0151 & 2 & 0.65 & 0.8443 & 0.180(2) & 0.738 \\
3.9 & 0.1310 & IMP  & 0.0200 & 2 & 0.66 & 0.8534 & 0.153(2) & 0.834 \\
3.9 & 0.1325 & IMP  & 0.0156 & 2 & 0.55 & 0.8540 & 0.146(2) & 0.702 \\
4.0 & 0.1301 & IMP  & 0.0200 & 2 & 0.66 & 0.8614 & 0.132(2) & 0.906 \\
4.0 & 0.1318 & IMP  & 0.0161 & 2 & 0.64 & 0.8625 & 0.121(2) & 0.799 \\
4.1 & 0.1283 & IMP  & 0.0200 & 2 & 0.75 & 0.8680 & 0.114(1) & 1.088 \\
4.1 & 0.1305 & IMP  & 0.0166 & 2 & 0.70 & 0.8685 & 0.104(1) & 0.668 \\
4.2 & 0.1246 & IMP  & 0.0200 & 2 & 0.86 & 0.8736 & 0.107(1) & 1.496 \\
4.2 & 0.1266 & IMP  & 0.0200 & 2 & 0.80 & 0.8738 & 0.097(1) & 1.346 \\
4.3 & 0.1253 & IMP  & 0.0200 & 2 & 0.83 & 0.8788 & 0.091(1) & 1.574 \\
4.4 & 0.1255 & IMP  & 0.0200 & 2 & 0.88 & 0.8836 & 0.086(1) & 1.411 \\
4.5 & 0.1253 & IMP  & 0.0200 & 2 & 0.83 & 0.8878 & 0.075(1) & 1.657 \\
4.6 & 0.1254 & IMP  & 0.0200 & 2 & 0.84 & 0.8916 & 0.072(1) & 1.617 \\
7.0 & 0.1315 & DBW2  & 0.0152 & 2 & 0.74 & 0.8344 & 0.252(6) & 0.780 \\
7.0 & 0.1345 & DBW2  & 0.0156 & 2 & 0.68 & 0.8352 & 0.233(8) & 0.673 \\
7.5 & 0.1310 & DBW2  & 0.0156 & 2 & 0.79 & 0.8516 & 0.206(3) & 0.779 \\
8.0 & 0.1305 & DBW2  & 0.0161 & 2 & 0.73 & 0.8663 & 0.168(2) & 0.764 \\
8.5 & 0.1300 & DBW2  & 0.0166 & 3 & 0.71 & 0.8774 & 0.134(1) & 0.782 \\
9.0 & 0.1224 & DBW2  & 0.0200 & 2 & 0.79 & 0.8858 & 0.137(3) & 1.412 \\
9.0 & 0.1296 & DBW2  & 0.0200 & 2 & 0.78 & 0.8865 & 0.115(1) & 0.753 \\
9.5 & 0.1228 & DBW2  & 0.0200 & 2 & 0.82 & 0.8934 & 0.109(2) & 1.576 \\
10.0 & 0.1234 & DBW2  & 0.0200 & 2 & 0.83 & 0.9000 & 0.099(2) & 1.502 \\
10.5 & 0.1236 & DBW2  & 0.0200 & 2 & 0.79 & 0.9056 & 0.093(1) & 1.567 \\
11.0 & 0.1239 & DBW2  & 0.0200 & 2 & 0.81 & 0.9110 & 0.086(1) & 1.473 \\
\end{tabular}
\end{center}
\caption{\label{tab:hmcresults} Simulation parameters and results for various dynamical simulations. The parameters are the gauge coupling, hopping parameter, gauge action, step size, and psuedofermion to gauge step size ratio. The results given are the mean link, lattice spacing (in fm, obtained from $r_0$ via the static quark potential) and pion mass (in GeV). Two degenerate flavours of FLIC sea quarks are used, with either L\"usher-Weisz (IMP) glue or DBW2 glue. These results are obtained from 20 $12^3 \times 24$ configurations. Simulations are done using multiple time step HMC with trajectories of unit length. Although an exact comparison is difficult, for a given step size and quark mass the acceptance rates obtained compare well with standard simulations (see for example Ref. \cite{cppacs-dynamical}).} 
\end{table*}

\end{document}